\newif\iffigs\figstrue

\documentstyle[12pt]{article}

 \stop \fi

\iffigs
  \input epsf
\else
\fi

\batchmode
  \newfont{\footscrfont}{rsfs10}
  \newfont{\footbbbfont}{msbm10}
\errorstopmode

\newif\ifscrf\scrftrue
\ifx\footscrfont\nullfont
  \scrffalse
\fi

\newif\ifamsf\amsftrue
\ifx\footbbbfont\nullfont
  \amsffalse
\fi

\def\ppnumber{\vbox{\baselineskip16pt\hbox{CLNS-95/1325}
\hbox{DUKE-TH-95-89}}}
\def\ppdate{March 1995}
\def\pplogo{\vbox{\kern-\headheight\kern -17pt
\halign{##&##\hfil\cr&{
\ppnumber}\cr\rule{0pt}{2.5ex}&\ppdate\cr}
}}

\makeatletter
\date{}
\def\dedicatory#1{\def\@date{\normalsize\it#1}}
\def\subjclass#1{\def\@thefnmark{}\@footnotetext{1991
    {\it Mathematics Subject Classification.} #1}}
\def\keywords#1{\def\@thefnmark{}\@footnotetext{
    {\it Key words and phrases.} #1}}

\def\ps@firstpage{\ps@empty \def\@oddhead{\hss\pplogo}%
  \let\@evenhead\@oddhead 
}
\def\maketitle{\par
 \begingroup
 \def\thefootnote{\fnsymbol{footnote}}
 \def\@makefnmark{\hbox
 to 0pt{$^{\@thefnmark}$\hss}}
 \if@twocolumn
 \twocolumn[\@maketitle]
 \else \newpage
 \global\@topnum\z@ \@maketitle \fi\thispagestyle{firstpage}\@thanks
 \endgroup
 \setcounter{footnote}{0}
 \let\maketitle\relax
 \let\@maketitle\relax
 \gdef\@thanks{}\gdef\@author{}\gdef\@title{}\let\thanks\relax}

\def\abstract{\if@twocolumn
\section*{Abstract}
\else \small
\begin{center}
{\bf ABSTRACT}
\end{center}
\quotation
\fi}

\newif\iffn\fnfalse

\@ifundefined{reset@font}{\let\reset@font\empty}{} 
\long\def\@footnotetext#1{\insert\footins{\reset@font\footnotesize
    \interlinepenalty\interfootnotelinepenalty
    \splittopskip\footnotesep
    \splitmaxdepth \dp\strutbox \floatingpenalty \@MM
    \hsize\columnwidth \@parboxrestore
   \edef\@currentlabel{\csname p@footnote\endcsname\@thefnmark}\@makefntext
    {\rule{\z@}{\footnotesep}\ignorespaces
      \fntrue#1\fnfalse\strut}}}
\makeatother

\ifamsf
  \newfont{\bigbbbfont}{msbm10 scaled\magstep2}
  \newfont{\bbbfont}{msbm10 scaled\magstep1}  
  \newfont{\smallbbbfont}{msbm8}
  \newfont{\tinybbbfont}{msbm6}
  \newfont{\smallfootbbbfont}{msbm7}
  \newfont{\tinyfootbbbfont}{msbm5}
\fi

\ifscrf
  \newfont{\scrfont}{rsfs10 scaled\magstep1}  
  \newfont{\smallscrfont}{rsfs7}
  \newfont{\tinyscrfont}{rsfs7}
  \newfont{\smallfootscrfont}{rsfs7}
  \newfont{\tinyfootscrfont}{rsfs7}
\fi

\ifamsf
  \newcommand{\Bbb}[1]{\iffn
      \mathchoice{\mbox{\footbbbfont #1}}{\mbox{\footbbbfont #1}}
      {\mbox{\smallfootbbbfont #1}}{\mbox{\tinyfootbbbfont #1}}\else
      \mathchoice{\mbox{\bbbfont #1}}{\mbox{\bbbfont #1}}
      {\mbox{\smallbbbfont #1}}{\mbox{\tinybbbfont #1}}\fi}
\else
  \def\bigbbbfont{\bf}
  \def\Bbb{\bf}
\fi

\ifscrf
  \newcommand{\Scr}[1]{\iffn
    \mathchoice{\mbox{\footscrfont #1}}{\mbox{\footscrfont #1}}
    {\mbox{\smallfootscrfont #1}}{\mbox{\tinyfootscrfont #1}}\else
    \mathchoice{\mbox{\scrfont #1}}{\mbox{\scrfont #1}}
    {\mbox{\smallscrfont #1}}{\mbox{\tinyscrfont #1}}\fi}
\else
  \def\Scr{\cal}
\fi

\def\operatorname#1{\mathop{\rm #1}\nolimits}
\def\C{{\Bbb C}}

\def\O{{\cal O}}
\def\P{{\Bbb P}}
\def\Q{{\Bbb Q}}

\def\Z{{\Bbb Z}}

\def\Hom{\operatorname{Hom}}

\def\Tors{\operatorname{Tors}}
\def\Free{\operatorname{Free}}

\def\opeq#1{\advance\lineskip#1 \advance\baselineskip#1
	\advance\lineskiplimit#1}

\def\eqalign#1{\null\,\vcenter{\opeq{2.5\jot}\mathsurround=0pt
	\everycr={}\tabskip=0pt
	\halign{\strut\hfil$\displaystyle{##}$&$\displaystyle{{}##}$\hfil
	\crcr#1\crcr}}\,\null}
\def\mapright#1{\smash{\mathop{\longrightarrow}\limits^{#1}}}
\def\mapdown#1{\Big\downarrow\rlap{$\vcenter{\hbox{$\scriptstyle#1$}}$}}

\def\sm{$\sigma$-model}

\def\CY{Calabi-Yau}

\def\cM{{\Scr M}}

\def\cD{{\Scr D}}

\def\cMc{{\hfuzz=100cm\hbox to 0pt{$\;\overline{\phantom{X}}$}\cM}}
\def\barcD{{\hfuzz=100cm\hbox to 0pt{$\;\overline{\phantom{X}}$}\cD}}

\def\ff#1#2{{\textstyle\frac{#1}{#2}}}

\makeatletter
\relax
\@writefile{toc}{\string\contentsline\space {section}{\string\numberline\space
{1}Introduction}{1}}
\@writefile{toc}{\string\contentsline\space {section}{\string\numberline\space
{2}The A-model}{3}}
\@writefile{lof}{\string\contentsline\space {figure}{\string\numberline\space
{1}{\ignorespaces The A-model moduli space for target space with ${\mathchoice
{\unhbox \voidb@x \hbox {\bbbfont Z}}{\unhbox \voidb@x \hbox {\bbbfont Z}}
{\unhbox \voidb@x \hbox {\smallbbbfont Z}}{\unhbox \voidb@x \hbox {\tinybbbfont
Z}}}_3\subset H_2(X)$.}}{5}}
\@writefile{toc}{\string\contentsline\space {section}{\string\numberline\space
{3}Away from Criticality}{6}}
\@writefile{lof}{\string\contentsline\space {figure}{\string\numberline\space
{2}{\ignorespaces The infra-red limit of the quintic threefold with 5 points
blown up.}}{10}}
\@writefile{toc}{\string\contentsline\space {section}{\string\numberline\space
{4}Examples}{11}}
\@writefile{toc}{\string\contentsline\space
{subsection}{\string\numberline\space {4.1}A Double Cover of $\hbox
{\bigbbbfont P}^3$}{11}}
\@writefile{lof}{\string\contentsline\space {figure}{\string\numberline\space
{3}{\ignorespaces The relationship between $X$, $X^\prime $ and $\mathaccent
"707E X$.}}{12}}
\@writefile{toc}{\string\contentsline\space
{subsection}{\string\numberline\space {4.2}A Double Cover of $(\hbox
{\bigbbbfont P}^1)^3$}{13}}
\@writefile{toc}{\string\contentsline\space {section}{\string\numberline\space
{5}Discussion}{15}}
\makeatother

\begin{document}
\setcounter{page}0
\title{\LARGE Stable Singularities in String Theory}
\author{
Paul S. Aspinwall\\[0.7cm]
\normalsize F.R.~Newman Lab.~of Nuclear Studies,\\
\normalsize Cornell University,\\
\normalsize Ithaca, NY 14853\\[10mm]
David R. Morrison\\[0.7cm]
\normalsize Department of Mathematics, \\
\normalsize Box 90320, \\
\normalsize Duke University, \\
\normalsize Durham, NC 27708-0320\\[10mm]
(with an appendix by Mark Gross)
}

{\hfuzz=10cm\maketitle}

\def\Large{\large}
\def\LARGE{\large\bf}

\vskip 1cm

\begin{abstract}

We study a topological obstruction of a very stringy nature concerned with
deforming the target space of an $N=2$ non-linear \sm. This target
space has a singularity which may be smoothed away according to the
conventional rules of geometry but when one studies the associated
conformal field theory one sees that such a deformation is not
possible without a discontinuous change in some of the correlation
functions. This obstruction appears to come from torsion in the
homology of the target space (which is seen by deforming the theory by
an irrelevant operator). We discuss the link between this phenomenon and
orbifolds with discrete torsion as studied by Vafa and Witten.

\end{abstract}

\vfil\break

\section{Introduction}		\label{s:intro}

A very interesting aspect of string theory is the way in which
space-time is described. In physics, thanks to the success of general
relativity, we are accustomed to picturing space-time as being a manifold
equipped with a metric. The physics of space-time is then described in
terms of this metric. Such a picture has some potential
shortcomings. In particular we may wish to consider some space-time
which is not smooth and thus may not admit a metric in the
conventional sense. One way
to treat such a space may be as a limit of a sequence of smooth
manifolds which converges to the desired space. Thus the ``metric'' on the
singular space is approximated by this sequence of smooth metrics.

While such a picture appears natural from a viewpoint of general
relativity it may be that it is not so natural from a string theory
point of view. In this paper we illustrate precisely this point by
considering a singular
space which classically appears as the limit of a sequence of smooth
manifolds and then showing that a string theory on the singular space
cannot be deformed into a string theory on any of the smooth manifolds
which approximate it.

The framework in which we will work is one of the most successful for
studying stringy aspects of geometry. That is, we look at $N$=(2,2)
superconformal field theories and their associated \CY\ target
spaces. We also restrict ourselves in this paper to the rich class of
complex dimension three target spaces. The usefulness of $N$=(2,2)
theories is that one can understand the geometry of the target \CY\
manifold without any explicit reference to the target space metric
(see, for example, \cite{me:N2lect} for a review). One may also study
many singular spaces, such as orbifolds \cite{DHVW:}, without any inherent
difficulties.

The deformations of a \CY\ manifold can be understood in terms of
marginal operators in the associated conformal field theory. If the
target space is singular, rather than being a manifold, the marginal
operators presumably still tell one how to deform the singular space. Certain
of these deformations may remove some, or perhaps all, of the
singularities. This process is well-understood in many
cases of orbifolds (see, for example, \cite{me:orb2}) where twisted marginal
operators in the conformal field theory can be matched to the
``blow-ups'' of the orbifold, i.e., deformations which resolve (at
least partially) the quotient singularities of the orbifold.

In \cite{VW:tor} some examples of more troublesome orbifolds were
studied. It was found that certain of the deformations of the classical
orbifold appeared to be ``missing'' in the conformal field theory
language. That is, these geometric deformations could not be seen by the string
theory. The purpose of this paper is to shed some light on the
geometrical explanation for such a phenomenon. We will see that there
is a truly stringy explanation for such obstructions. These
obstructions are due to world-sheet instantons wrapping themselves
around particular elements of the second homology group of the target
space.

The construction of \cite{VW:tor} rests upon the study of ``discrete
torsion.'' There is some potential for confusion on the
subject of discrete torsion and we will clearly set out our
definitions here. Given a conformal field theory for a \CY\ manifold
$V$ with a discrete symmetry group $G$, one may build the theory for
the quotient $V/G$ in a systematic way. There is an ambiguity in this
construction however. Phases may be introduced when building the
partition function for the characters without disturbing modular
invariance. It was shown in \cite{Vafa:tor} that these phases must be
elements of $H^2(G,U(1))$.
If $G$ is finite then this group is isomorphic to $H_2(G)$. (The coefficient
group
$\Z$ is assumed for homology and cohomology if omitted.) Thus each
element $\varepsilon\in H^2(G,U(1))$ gives rise to a possible
conformal field theory for the orbifold $V/G$. We call $\varepsilon$
the ``2-cocycle'' for the theory. In \cite{VW:tor} some examples of
orbifolds with a nontrivial 2-cocycle
were studied and it was shown that each of the marginal operators
could be associated to a deformation of the orbifold space
itself, but that some deformations appeared to be ``missing,'' i.e.,
corresponded to no marginal operator.

The singular cohomology groups $H^*(X)$ of a manifold $X$ need not be
free abelian groups. Of particular interest to us in this paper will
be the torsion part of $H_2(X)$ (or equivalently $H^3(X)$) where $X$
is a \CY\ manifold. We impose the condition
$h^{2,0}(X)=0$. In this case the torsion group is isomorphic to the ``Brauer
group'' of $X$. We will use this terminology here for convenience
although the reader is not required to know the full definition of the
Brauer group.\footnote{Further information about the Brauer group is
provided in the appendix.}
 It was suggested in \cite{Vafa:tor} that there should be
some connection between the Brauer group and the 2-cocycles in an
orbifold. An example studied in \cite{AM:suff} had trivial Brauer
group but admitted nontrivial 2-cocycles. Thus these two concepts are
not equivalent. As we shall see in this paper however there is some
intimate connection between them. Because the term ``discrete
torsion'' has been used at times to refer to either the Brauer group
or the group of $2$-cocycles,
we will try to avoid using it in this paper to save
any confusion.

Although motivated by the orbifold construction of \cite{VW:tor} we
shall see that stable singularities are probably not confined to such
examples. The topological obstruction to deforming away the
singularities may be thought of as ``hiding'' away in the singularity
itself. To understand this geometrically we will blow up the
singularity to expose its contents. This blow-up will not be a
marginal perturbation as one is accustomed to in orbifold theory but
rather will be an irrelevant perturbation.

In section \ref{s:Amod} we discuss how the Brauer group affects the
correlation functions of an $N$=(2,2) superconformal field theory. In
particular we only need concern ourselves with that part of the
conformal field theory which is present in the A-model (which is one
of the topological field theories obtained by twisting the original
$N$=2 model). We will review how the Brauer group adds a degree of
freedom to the A-model that cannot be expressed in terms of the
K\"ahler form or the $B$-field.

In section \ref{s:ncrit} we discuss blow-ups as irrelevant
operators. This generalizes the usual notion of blow-ups in the
context of orbifold theories which correspond to truly marginal operators.
We will need such a generalization to deal with the singularities
discussed in this paper.
This allows us to study the examples of stable
singularities in section \ref{s:egs}.

Finally we present a discussion
in section \ref{s:conc}.


\section{The A-model}		\label{s:Amod}

In this section we will study the form of the correlation functions of
the the A-model with target space $X$ where $X$ may have a non-trivial
Brauer group, that is, when $H^3(X)$ contains a torsion subgroup. From
the universal coefficient theorem (see, for example, \cite{BT:}) the
torsion part of $H^3(X)$ is isomorphic to the torsion part of $H_2(X)$.

The A-model is a topological field theory \cite{W:AB} in which the
correlation functions depend upon non-trivial instanton effects. The
instantons are holomorphic maps from the world-sheet, $\Sigma$, to the target
space $X$. Further, the action of this instanton is assumed to depend
only upon the homology class of the image of this map in $X$. For an
instanton $I$ with homology class $[I]\in H_2(X)$, let us denote
$e^{-S_I}$ by $\mu([I])$, where $S_I$ is the action of the
instanton. In order for string interactions to behave correctly
\cite{Vafa:tor} we further demand that the action depend linearly upon
the homology class of $I$, i.e.,
\begin{equation}
  \mu\in\Hom(H_2(X),\C^*),
\end{equation}
where $\C^*$ is the multiplicative group of nonzero complex numbers.

Recall that $H_2(X)$ is an abelian group and thus may be decomposed
into free and torsion subgroups:
\begin{equation}
  H_2(X) \cong \Z^{h^{1,1}(X)} \times \Z_{t_1}\times
	\Z_{t_2}\times\ldots,
\end{equation}
where $t_i$ are finite positive integers labeling the torsion part of
$H_2(X)$.

A simple application of the universal coefficient theorem
tells us that $\Hom(H_2(X),\C^*)\cong H^2(X,\C^*)$.
Given the exact sequence
\begin{equation}
  0\to\Z\to\C\to\C^*\to0,
\end{equation}
we obtain the long exact sequence
\begin{equation}
  0\to\Free(H^2(X))\to H^2(X,\C)\to H^2(X,\C^*)\to \Tors(H^3(X))\to0,
		\label{eq:les1}
\end{equation}
where $\Free(G)$ and $\Tors(G)$ denote the free and torsion parts of
the abelian group $G$ respectively.

For the time being let us assume that $\Tors(H^3(X))\cong0$. Then we
see from (\ref{eq:les1}) that
\begin{equation}
  H^2(X,\C^*)\cong\frac{H^2(X,\C)}{\Free(H^2(X))}.
	\label{eq:noBr}
\end{equation}
We usually think of the A-model correlation functions as depending
upon the ``complexified K\"ahler form.'' This complexified K\"ahler
form is written $B+iJ$ where $J$ is the usual K\"ahler form and $B$ is
a real 2-form of $X$ defined modulo elements of de Rham cohomology
which are elements of integral cohomology. It is easy to see that
this agrees with (\ref{eq:noBr}).

Let us form a basis of $\Free(H^2(X))$ with elements $e_j, j=1\ldots
h^{1,1}(X)$. We may then expand
\begin{equation}
  B+iJ=\sum_j(B+iJ)_j e_j,
\end{equation}
where $B_j\cong B_j+1$ for all $j$. We now introduce the usual
$q$-variables:
\begin{equation}
  q_j=\exp\left\{2\pi i(B+iJ)_j\right\}.
\end{equation}
We then see that
\begin{equation}
  \mu([I])=\prod_j q_j^{n_j([I])},
\end{equation}
where $n_j$ are non-negative integers labeling the homology class of
the instanton.

When we calculate a correlation function in the A-model we use the
usual methods of intersection theory in topological field theory
\cite{W:aspects}. That is to say, for each instanton background, the
contribution to the correlation function is given by the intersection
number of some cycles representing the observables in the moduli space
of the instanton. This intersection number is an
integer\footnote{It is {\it a priori}\/ possible that the instanton
moduli spaces in some examples could
have orbifold singularities, leading to rational numbers rather than
integers. No examples of this phenomenon are known, and perhaps it
does not occur.  In any case, although we have assumed here that the
intersection numbers are integers, the
discussions in this paper are unchanged if rational numbers are
used in place of integers.}. Such a
contribution to the correlation function is then weighted by
$\mu([I])$. We thus see that the correlation functions in this
A-model take the form of power series in the variables $q_j$ with
{\it integer\/} coefficients. All of the many examples studied so far
confirm this (see, for example \cite{CDGP:,CDFKM:I}).

Now let us consider the case when the Brauer group of $X$ is not
trivial. We take the simplest case where
$\Tors(H^3(X))\cong\Z_m$ for some integer $m$. This implies that
$\Tors(H_2(X))\cong\Z_m$. Let $t$ be a 2-cycle so that $[t]$ generates
this torsion class. That is, $t$ is a cycle such that
\begin{equation}
  p[t]\cong0 \:\Leftrightarrow\: m \mathbin{\rm divides} p.
\end{equation}
This implies that $(\mu([t]))^m=1$, i.e.,
\begin{equation}
  \mu([t]) = \exp\left(\frac{2\pi i}m\alpha\right),
\end{equation}
for $\alpha=0,\ldots,m-1$.

The choice of $\alpha$ is required, in addition to $B+iJ$, to
determine the correlation functions. That is, the position of the
specific A-model in the moduli space of theories is not entirely
determined by the complexified
K\"ahler form but also requires a specification of the discrete
parameter $\alpha$.

We may consider the shape of the moduli space of A-models as
follows. The ``large radius limit'' of an A-model is the limit point
where the action of all nontrivial instantons becomes infinite. That is,
$\mu([I]) \to 0$ for $[I]\not\cong0$ (as always we have $\mu(0)=1$). At
this limit point therefore the
choice of $\alpha$
does not matter. Thus different ``sheets'' of the moduli space, each
parametrized by $B+iJ$ but having a different value of $\alpha$, are
joined at the large radius limit. It is important to remember that
when describing the moduli space in terms of A-models we assume that
we are in the neighbourhood of the large radius limit and that each
correlation function is completely determined by a power series centered
at the limit point. Thus, we will
not discuss further aspects of the global geometry of the moduli
space which take us outside this region. This region of the moduli
space is shown in figure \ref{fig:Amod}.

\iffigs
\begin{figure}
  \centerline{\epsfxsize=13cm\epsfbox{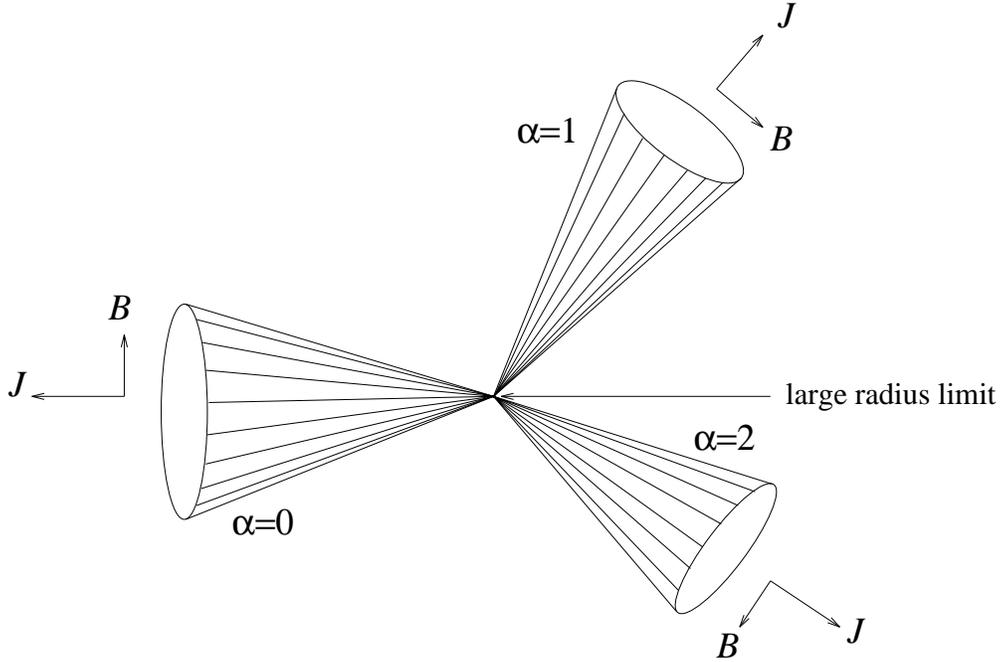}}
  \caption{The A-model moduli space for target space with $\Z_3\subset
H_2(X)$.}
  \label{fig:Amod}
\end{figure}
\fi

Let us illustrate the new form of the correlation function by an
example in which $m=3$ and we are considering observables corresponding
to divisors in $X$ where $X$ is a \CY\ threefold. For simplicity let
us also assume that $h^{1,1}(X)=1$ so that there is only one such observable.
Thus we have $H_2(X)\cong\Z\times\Z_3$. Now we may perform the usual
expansion of the A-model correlation functions in terms of the
rational curves on $X$ as was done in \cite{CDGP:}. In
this case we expect
\begin{equation}
  \langle \O_D\O_D\O_D\rangle=\#(D\cap D\cap D)+(n_1+\omega^\alpha n_2
	+\omega^{2\alpha}n_3)q+O(q^2),	\label{eq:D3}
\end{equation}
where $\omega$ is a nontrivial cube root of unity.
It is not possible for any algebraic curve to lie in a torsion class of
$H_2(X)$. This is because the area of an algebraic curve is given by
the integral of the K\"ahler form over the curve. This area may be
recast as the intersection of a 4-cycle representing the dual of the
K\"ahler form with the curve. Since any cycle in a torsion class must
have zero intersection number with any other cycle, any curve in a
torsion class would have zero area, which is not possible.
It may happen however that the difference of two curves is a torsion cycle.
The lines (i.e., rational curves intersecting $D$ once) on $X$ can
therefore lie in one of three homology classes and these are counted
by $n_1,n_2,n_3$. The number of lines is thus $n_1+n_2+n_3$ which is
counted by (\ref{eq:D3}) in the usual way when $\alpha=0$. When
$\alpha$ is 1 or 2 however we distinguish between these lines.

Note that for $\alpha=1$ or 2 the series (\ref{eq:D3}) is {\em not\/} a power
series with integer (or rational) coefficients. Thus the fact that all examples
studied so far did lead to a series with integer coefficients implies
that the examples had a trivial Brauer group (or at least, only
elements of order 2). It would be interesting to study an example of a
smooth \CY\ manifold with non-trivial Brauer group and so generate a
solid example of the series of the form (\ref{eq:D3}). Unfortunately
at this point in time we are not aware of any such examples.


\section{Away from Criticality}		\label{s:ncrit}

As is well-known (see, for example, \cite{GSW:book}) the
$\beta$-function for the metric of a non-linear \sm\ is given by the
Ricci-tensor to first order
\begin{equation}
  \beta(g_{ij}) = -\ff1{2\pi}R_{ij}+\ldots
\end{equation}
If the target space is at large radius, the later terms in the series
are negligible. Consider the flow towards the infra-red (low-energy)
limit. If the target space has positive curvature then the sign of the
$\beta$-function shows that the space will
will shrink under this flow.
If the space has negative curvature it will expand, and if
it is Ricci-flat then it will be stable (to leading order). \CY\
manifolds fall into the last category and thus may provide conformally
invariant \sm s.

A complex projective space is a space of positive curvature. The
non-linear \sm\ on such a space is a massive field theory \cite{ADL:CPn} and
thus na\"\i vely appears to flow to something trivial in the infra-red
limit. This may be
viewed geometrically as a process in which
 the target space shrinks down to a point in the
limit. Actually one needs to be a little careful about this
statement. Although one might think that a non-linear \sm\ with a
point target space is, by its very definition, a trivial theory, one
may reach different conclusions by treating the point as a limit point of
theories on a complex projective space \cite{CV:CPn}. In the latter
case the limit of the infra-red flow is better thought of as a target
space whose size is $-\infty$. This has become a recurring theme in
recent works \cite{W:phase} and may be viewed as part of the inherent
difficulty in clearly defining the concept of sizes below the Planck
scale.

The best method of giving geometrical interpretations to spaces away
{}from the large radius limit is probably that of the linear \sm\ of
\cite{W:phase}. The behaviour of the complex projective space, $\P^n$, as a
target space was studied in the latter and gave the following picture
which is most complete version of what happens in the infra-red limit.
The linear \sm\ contains a real parameter, $r$, which, in the case
$r\gg0$ gives the size of the complex projective target space (or, to
be more precise, the area of complex lines in the target space). The
renormalization group acts on this parameter to drive it towards
$-\infty$ in the I.R.~limit. In this limit however the geometrical
interpretation changes. When $r\ll0$ the target space becomes that of
$(n+1)$ disjoint points. The conformal field theory associated with such a
target space has $c=0$ but consists of $(n+1)$ $Sl(2,\C)$-invariant vacua.
That is, we have a reducible but trivial representation of the
Virasoro algebra. This is the sense in which the $\P^n$-model flows to
a trivial theory in the I.R.\ limit. Note that this picture preserves
the Witten index, $\operatorname{Tr}(-1)^F$, or Euler characteristic,
of the theory during the
flow. The Euler characteristic of both $\P^n$ and $(n+1)$ disjoint
points is $n+1$.

\def\bX{\tilde X}
We wish now to consider something intermediate between a projective
space and a \CY\ manifold. That is, we want a theory which is not
conformally invariant but flows to a non-trivial conformal field
theory in the infrared limit. Such an example may be provided by
blowing-up a smooth point on a \CY\ manifold. Blowing up points is
familiar in string theory for resolving orbifold singularities (see
\cite{me:orb2} for a review). Blowing up singularities may result in a
smooth \CY\ manifold. If one blows up a point on a \CY\ manifold, $X$,
that
is already smooth however one obtains a manifold, $\bX$, which does not admit
a Ricci-flat metric (although it is still complex and K\"ahler).

In the case of a \CY\ threefold, blowing up a smooth point replaces
that point by a divisor isomorphic to
the projective space $\P^2$. The normal bundle of this
divisor is $\O(-1)$ (i.e., the inverse of the Hopf bundle). Consider a
curve $C$ which is a projective space $\P^1$ lying within this
$\P^2$. It is a simple matter to show that the normal bundle of the
curve is $\O(1)\oplus\O(-1)$. Given that its tangent bundle is $\O(2)$
we obtain
\begin{equation}
  \int_C c_1(\bX) = 2+1-1 = 2.
\end{equation}
Thus $c_1\neq0$. In particular since $\int_Cc_1(\bX)>0$, the curve $C$
will shrink during the flow to the infra-red limit. Since all such curves
shrink, the ``exceptional divisor'' $\P^2$ will also shrink. Curves
away from this blowup will satisfy $\int_Cc_1(\bX)=0$ and should be
stable under this flow. So long as a neighbourhood of such a curve is
stable under the flow, the process must lead to a birational
transformation and so the complex structure remains fixed.
Thus, the net result would appear to be that
the limit of this flow is to turn $\bX$ back into $X$. That is, we
take a manifold $X$ corresponding to a conformal field theory. We then
perturb it to obtain a field theory that is not conformally invariant,
but flows back to the original under flow to the infra-red limit.
In other words,
{\it blowing up a smooth point is equivalent to perturbation by an
irrelevant operator}.

It is worth describing an example of this picture in terms of the
linear \sm, as we now do for completeness. The reader who is already convinced
of our assertions concerning the effects of the renormalization group
may skip this section. Let us consider the case of the quintic
hypersurface in $\P^4$. This is a smooth \CY\ manifold. A generic line
(i.e., linearly embedded $\P^1$) in the ambient $\P^4$ will intersect
this \CY\ manifold at five distinct points. Thus by blowing up such a
line we blow up the \CY\ manifold at five points. The toric picture of
this blown-up ambient space leads to the following gauged linear \sm.
Consider seven chiral superfields with lowest components
$x_1,\ldots,x_5,p,f$ in a theory with gauge group $U(1)^2$. The
charges are as follows:
\begin{equation}
\begin{tabular}{|c|c|c|}
  \hline
  &$Q^{(1)}_i$&$Q^{(2)}_i$\\
  \hline
  $x_1$&1&1\\
  $x_2$&1&1\\
  $x_3$&1&1\\
  $x_4$&1&0\\
  $x_5$&1&0\\
  $p$&$-5$&0\\
  $f$&0&$-1$\\
  \hline
\end{tabular}
\end{equation}
Part of the classical potential comes from the $D$-terms of this
theory and the vanishing of this requires that two parameters of the
theory $r_1$ and
$r_2$ be set as follows
\begin{equation}
  \eqalign{r_1 &= |x_1|^2+|x_2|^2+|x_3|^2+|x_4|^2+|x_5|^2-5|p|^2\cr
    r_2 &= |x_1|^2+|x_2|^2+|x_3|^2-|f|^2.\cr}
\end{equation}
We also consider the invariant superpotential
\begin{equation}
  W = p\left(f^5(x_1^5+x_2^5+x_3^5) + x_4^5 + x_5^5\right),
                                             \label{eq:W}
\end{equation}
as in \cite{W:phase}. The vanishing of the classical potential
requires all of the derivatives of (\ref{eq:W}) to be zero.

Now consider the phase where $r_1-r_2>0$ and $r_2>0$. With this choice,
the classical vacuum requires that at least one of $\{x_1,x_2,x_3\}$
does not vanish and that at least one of $\{x_4,x_5,f\}$ does not
vanish either. Suppose first that $f\neq0$. We may fix the phase of $f$
(we then normalize $f=1$)
using one of the $U(1)$ groups. The derivatives of the superpotential
then require $p=0$ and that
\begin{equation}
  x_1^5+x_2^5+x_3^5+x_4^5+x_5^5=0.
\end{equation}
Let the other $U(1)$ action be used to form $[x_1,x_2,x_3,x_4,x_5]$ as
the homogeneous coordinates of $\P^4$ (expressing $\P^4$ in the
familiar form of a symplectic
reduction $S^9/U(1)$). Thus the classical vacuum appears to be the
quintic \CY\ hypersurface in $\P^4$. Note however that we are missing
the line $[0,0,0,x_4,x_5]$ and hence 5 points of this \CY\ manifold.
Now let $f=0$. This forces either $x_4$ or $x_5$ to be nonzero and
thus $p=0$. We also have the constraint $x_4^5+x_5^5=0$ from one of
the derivatives of the superpotential. Use one of the $U(1)$'s to fix
the phases of $x_4$ and $x_5$. The other $U(1)$ may be used to form
$\P^2$ with homogeneous coordinates $[x_1,x_2,x_3]$. The result is
that each of the 5 points $[0,0,0,x_4,x_5]$ in the quintic
hypersurface have been replaced by $\P^2$. That is, we have blown-up 5
smooth points as promised.

Next consider the phase where $r_1>0$ and $r_2<0$. Now $f$ must be
nonzero and we fix it using one of the $U(1)$ groups. Also one of
$\{x_1,x_2,x_3,x_4,x_5\}$ are nonzero and we use the other $U(1)$
to form $\P^4$. It follows that $p=0$ and we lie on the quintic
hypersurface. At first sight therefore, this phase appears to be simply
the quintic \CY\ manifold in $\P^4$. This is not the full story
however. Thus far we have neglected some of the fields in the theory
--- namely the lowest components of the twisted chiral superfields
coming from the field strength of the two $U(1)$ gauge fields. Call
these fields $\sigma_1$ and $\sigma_2$ consistent with the notation of
\cite{W:phase}. The classical potential of these fields is given by
\cite{MP:inst}:
\begin{equation}
  \eqalign{U_\sigma &= 2\sum_{a,b} \bar\sigma_a\sigma_b
    \sum_i Q^{(a)}_iQ^{(b)}_i |\phi_i|^2\cr
  &= 2|\sigma_1|^2(|x_1|^2+|x_2|^2+|x_3|^2+|x_4|^2+|x_4|^2+25|p|^2)\cr
     &\qquad+2(\sigma_1\bar\sigma_2+\bar\sigma_1\sigma_2)
      (|x_1|^2+|x_2|^2+|x_3|^2)\cr
     &\qquad\qquad+2|\sigma_2|^2(|x_1|^2+|x_2|^2+|x_3|^2+|f|^2).}
\end{equation}
It would appear that for $r_1\gg0$ and $r_2\ll0$ the fields $\sigma_a$
are very massive and so should be set equal to zero. It turns out
however that there are large quantum corrections to the potential when
$x_1=x_2=x_3=f=x_4^5+x_5^5=0$ and $\sigma_2$ appears to be massless.
One may show \cite{W:phase} that there are $\sum_i Q^{(2)}_i=2$ extra
solutions for $\sigma_2$ when $r_2\ll0$.

Our target space for this latter phase is thus as follows. We have a
smooth hypersurface in $\P^4$ where the $\sigma$ fields are zero and
we have a completely disjoint set of 10 points given by the 5 points
on the quintic with $x_1=x_2=x_3=0$ each with two possible nonzero
values for $\sigma_2$, and $\sigma_1$ is still zero. Note that the
Euler characteristic of this set is equal to $-200+10=-190$ which is
precisely that of the quintic blown-up at 5 points.

The effect of the I.R.\ flow is to force $r_2\to-\infty$. Thus if we
begin with a target space of the quintic threefold with 5 points
blown-up and go to the I.R.\ limit, we end up with the quintic
threefold with 10 disjoint points. This is shown in
figure~\ref{fig:blow}. Note that the resulting conformal field theory
consists of that of the quintic together with 10 trivial representations
of the Virasoro group. Thus it is only when we focus on the nontrivial
irreducible part of the conformal field theory that the blow-up is,
strictly speaking, an irrelevant operator. Since our main concern in
this paper is the behaviour of the correlation functions only in this
part of the theory, this meaning of an irrelevant operator is good
enough for our purposes. Note also that the complex structure on the
target space is unaffected by this process as expected.

\iffigs
\begin{figure}
  \centerline{\epsfxsize=13cm\epsfbox{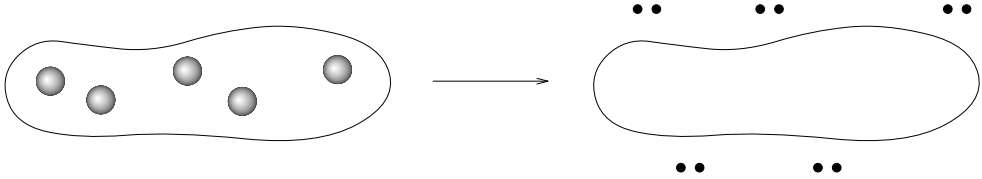}}
  \caption{The infra-red limit of the quintic threefold with 5 points
blown up.}
  \label{fig:blow}
\end{figure}
\fi

Let us briefly note that the ``monomial-divisor mirror map'' of
\cite{AGM:II,AGM:mdmm} may also be extended to this picture. This
map maps a toric divisor in the target space of the A-model to a
monomial which may be used to deform the complex structure of the
mirror B-model. This is done by identifying both the sets of divisors
and the set of monomials with a set of points lying in a hyperplane
intersecting a lattice based on the ideas of \cite{Bat:m}. The blow-up
of a smooth point can be
represented by a point outside this hyperplane, this in turn maps to a
monomial with the wrong weight to be considered as part of the
original quasi-homogeneous defining equation. In fact, the monomial's
weight is too high and thus is of no importance in the infrared limit
as was argued in \cite{VW:}. Thus we see again that the blow-up is an
irrelevant operator. Presumably given a good definition of the B-model
away from criticality, we could understand the meaning of the trivial
representations that appear in the I.R.\ limit.

Thus far we have gained little. We already knew how to handle a
non-linear \sm\ on a smooth \CY. Where blow-ups prove useful is where
they resolve singularities. If the target space $X$ is singular then
one may study the \sm\ by blowing $X$ up. Depending on the first Chern
class of the resulting target space this may be a relevant, marginal
or irrelevant perturbation of the original theory. In the case of
orbifolds, the cases considered are usually marginal (see, for
example, \cite{me:orb2}). One may also have a case of a singularity being
resolved by an irrelevant operator as we now show.

Consider $\C^4$ with coordinates $(w,x,y,z)$ and the hypersurface
defined by the equation
\begin{equation}
  xy=wz.		\label{eq:doublept}
\end{equation}
This hypersurface has an isolated singularity, or ``node,'' at the
origin. There are
many ways to remove such a singularity. Consider a compact variety $X$
which contains such a node. The ways in which $X$ can be
smoothed depend upon the global geometry of $X$. That is, there may be
global obstructions to processes which removed the singularity
locally. Whether or not the resulting smooth space is K\"ahler is also
a global question.

If $X$ is a projective algebraic variety then there is always at least one way
of smoothing $X$ to form a K\"ahler manifold, $\bX$, as follows.
Blow up the origin of $\C^4$ in which the node (\ref{eq:doublept}) is
embedded. Thus, the origin is replaced by $\P^3$. The intersection of
the hypersurface (\ref{eq:doublept}) with this $\P^3$ is obtained by
treating the coordinates $(w,x,y,z)$ as homogeneous coordinates. Thus
the effect of the blow-up is to replace the node by a quadric
hypersurface in $\P^3$. It is a well-known result in algebraic
geometry that such a complex 2-fold is isomorphic to $\P^1\times\P^1$.

Now consider the first Chern class of this blow-up. Let $C$ by any one
of the two families of rational curves in the exceptional divisor
$\P^1\times\P^1$. Clearly because of the product structure, the normal
bundle of this curve within the exceptional divisor is $\O(0)$. The
other normal direction of the curve is that of the way $\P^3$ embeds
in $\C^4$. Thus the total normal bundle of the curve is
$\O(0)\oplus\O(-1)$. Adding this to the tangent bundle we obtain
\begin{equation}
  \int_C c_1(\bX) = 2+0-1 = 1.
\end{equation}
This is thus similar to the case of a blow-up of a smooth
point --- the blow-up is an irrelevant operator. In contrast to the
latter case however, the irrelevant perturbation has been of
some use --- we have smoothed the target space.


\section{Examples}	\label{s:egs}

We are now in a position to study some examples for the target space
which exhibit
stable singularities thanks to the analysis of the preceding
sections.

\subsection{A Double Cover of $\hbox{\bigbbbfont P}^3$}
		\label{ss:eg1}

Let $K^\prime$ be a smooth hypersurface in $\P^3$ defined by an equation of
degree 8 in the homogeneous coordinates. Let $X^\prime$ be a double
cover of this $\P^3$ branched over $K^\prime$. The space $X^\prime$ is
a smooth \CY\ manifold and was studied in the physics literature long
ago \cite{SW:coup}. The resulting space has $h^{1,1}=1$ given by the
original $\P^3$ and $h^{2,1}$=149 where the 149 corresponding
deformations of complex structure of $X^\prime$ can be provided by the 149
inequivalent deformations of the octic defining equation for
$K^\prime$. The Brauer group of $X^\prime$ is trivial.

By deforming the octic equation to special values we may make the
double cover singular. Let $X$ be such a singular degeneration of
$X^\prime$ branched over the singular octic surface $K$. We define $K$
as follows. Let $W$ be a generic polynomial in the homogeneous coordinates
$[x_0,\ldots,x_3],[p_0,\ldots,p_3]$ of degree (2,2) with respect to
the $x$'s and the $p$'s. The $x$'s form the homogeneous coordinates of
our $\P^3$. $K$ is defined by
\begin{equation}
  \left|\frac{\partial^2W}{\partial p_i\partial p_j}\right|=0.
\end{equation}
The number of nodes may be calculated using the methods of
\cite{HarTu:}\footnote{The matrix $\partial^2 W/\partial p_i\partial
p_j$ represents a symmetric map $f:E\to E^*$ where $E$ is a vector
bundle of rank 4 over $\P^3$. The fact that this matrix has entries which are
quadratic in the homogeneous coordinates of this $\P^3$ shows that
$E^*\cong\O(1)^{\oplus 4}$. Theorem 1 of \cite{HarTu:} can then be
used calculate the number of nodes since nodes appear as the locus of
corank $\ge2$ maps.}.
This surface has 80 nodes and as a result $X$ has 80 isolated nodes of
the form (\ref{eq:doublept}). $K$, and thus, $X$ have 69
deformations. The fact that $69+80=149$ shows that each of the nodes
appears to have ``eaten up'' one of the original deformations of
$X^\prime$.

Given $X$ how may we remove the singularities? Obviously one way is to
deform it back into $X^\prime$ which is a smooth \CY\
manifold. Another way one might be tempted to try is to use ``small
resolutions.'' This amounts to replacing each node by a $\P^1$. This
process was used in the string context in \cite{CGH:con} to continuously
change the topology of the target space. It is also reviewed in
\cite{me:N2lect} together with its relation to the ``flop.'' Anyway,
in this case the small resolutions don't work --- the resulting smooth
space is not K\"ahler.

Consider $\bX$ as the blow-up of $X$ replacing each of the 80 nodes by
exceptional divisors in the form of
$\P^1\times\P^1$ as described in the previous section. See figure
\ref{fig:eg1} (where only 2 of the 80 nodes are shown).
The space $\bX$
is smooth and K\"ahler but not \CY. As shown in the appendix however,
this space is very interesting for our purposes because $H_2(\bX)$
contains a $\Z_2$ subgroup. That is, we have an example with a
nontrivial Brauer group.

\iffigs
\begin{figure}
  \centerline{\epsfxsize=13cm\epsfbox{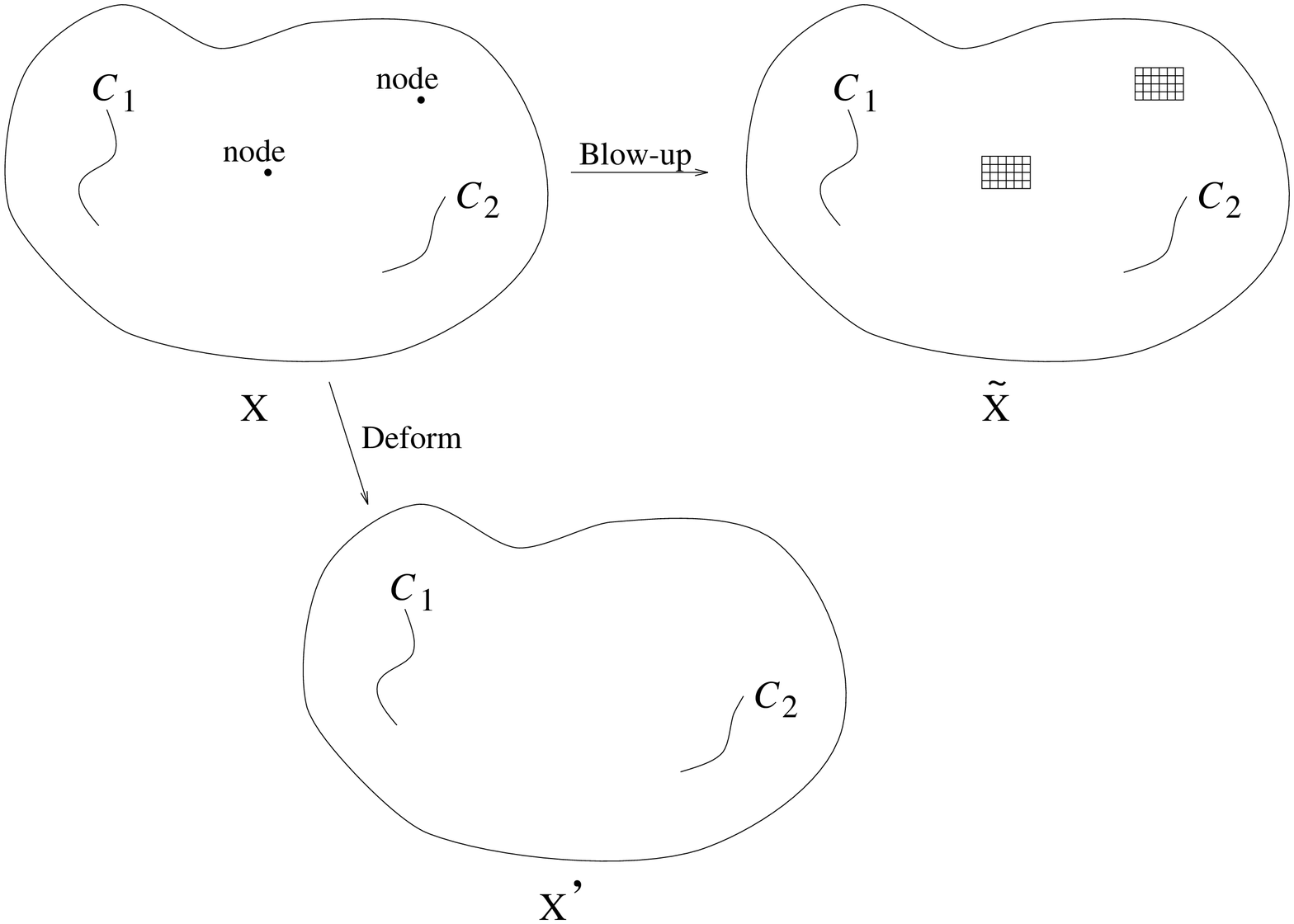}}
  \caption{The relationship between $X$, $X^\prime$ and $\bX$.}
  \label{fig:eg1}
\end{figure}
\fi

Now let us consider the A-model on $X$. Since $X$ is singular we need
to think carefully about how to calculate correlation functions in the
model. The $N=2$ \sm\ on $\bX$ flows to the superconformal field
theory on $X$ as explained in section \ref{s:ncrit}. Thus if the
exceptional divisors in $\bX$ are very
small then we expect to have a theory with correlation functions
very close to that of $X$. The exception to this will be correlation
functions involving fields from the part of the theory that became
trivial in the
infra-red limit. From section \ref{s:ncrit} we expect that the fields
associated with the homology of the
exceptional divisors themselves are such fields.  The A-model on $X$
can be considered as the limit of this infra-red flow, twisted to form a
topological field theory. The A-model on $X$ therefore would appear to
be given by the
A-model on $\bX$ with all the ``massive'' bits ignored since these
disappear in the infra-red limit. That is, we ignore the contributions
to the homology appearing from the exceptional divisors
themselves. What we do not ignore however is the torsion in $H_2(\bX)$
because this may be observed away from the exceptional divisors. In
particular there are rational curves $C_1,C_2\in\bX$ such that
$[C_1]-[C_2]$ is a nontrivial element of $\Tors(H_2(\bX))$ and neither
$C_1$ nor $C_2$ is contained in any of the exceptional divisors.

We propose therefore that one may define an A-model on $X$ in terms of
the homology classes on $\bX$ excluding classes lying exclusively
within the exceptional divisors. This means that our $\Z_2$ group in
the Brauer group allows us to introduce a parameter $\alpha$ which may
be 0 or 1 as in section \ref{s:Amod}.
In particular, if $\alpha=0$, the curves $C_1$ and $C_2$ will
contribute identically to correlation functions and if $\alpha=1$
they will contribute differently.

Consider this A-model as we deform $X$ slightly into $X^\prime$. All the
rational curves away from the nodes are deformed slightly but now the Brauer
group is trivial and so $C_1$ and $C_2$ lie in the same homology
class. In the underlying conformal field theory one would expect the
correlation functions to change slightly. This is all well and good if
$\alpha=0$ but if $\alpha=1$ then we are in trouble. The coefficients
in the A-model correlation functions
appear to jump as the homology classes of $C_1$ and
$C_2$ change.

Thus, an A-model on $X$ for which the parameter $\alpha$ is $0$ may possibly
be deformed into an A-model on $X'$, but for A-models with $\alpha=1$
this deformation appears to be
obstructed at the level of correlation functions. Since this
deformation is the
only way of smoothing $X$ into a \CY\ manifold, if $\alpha=1$ then we
are unable to follow the A-model from the singular space to the
smooth one.  Thus,
even though $X$ may be classically deformed into the \CY\ manifold
$X^\prime$, this deformation is not compatible with string theory!
The only deformations of complex structure of $X$ allowed
in the case $\alpha=1$ are the 69 which preserve the 80 nodes.

This state of affairs is, of course, similar to that suggested in
\cite{VW:tor} to which we now turn our attention.

\subsection{A Double Cover of $(\hbox{\bigbbbfont P}^1)^3$}
	\label{ss:eg2}

To discuss our next example,
we need to introduce a whole plethora of spaces all of which may be
deformed into each each other continuously. These are as follows:
\begin{itemize}
  \item[$X^\sharp$] Let $T$ be the torus of one complex dimension
described as a quotient of the complex plane $\C$, parametrized by $z$, with
identifications $z\cong z+1$ and $z\cong z+i$. Take three copies of
this torus parametrized by $z_1,z_2,z_3$. $X^\sharp$ is defined as the
orbifold obtained by dividing this space $T^3$ by the group
$G\cong\Z_2\times\Z_2$ generated by $(z_1,z_2,z_3)\mapsto(-z_1,-z_2,z_3)$
and $(z_1,z_2,z_3)\mapsto(z_1,-z_2,-z_3)$.
  \item[$Y$] This orbifold may be blown up in the usual way to form a
\CY\ manifold $Y$. $Y$ has $h^{1,1}=51$ and $h^{2,1}=3$. Actually this
blow-up is not unique and there are many topologies possible for
$Y$. Which one we choose is not important for this paper.
  \item[$X^\prime$] Consider the space $(\P^1)^3$ and a smooth hypersurface
$K^\prime$ within this space defined by an equation of weight
$(4,4,4)$ (i.e., quartic in each of three sets of homogeneous
coordinates). $X^\prime$ is the double cover of $(\P^1)^3$ branched
over $K^\prime$. It is a smooth \CY\ manifold with torsion-free
cohomology with $h^{1,1}=3$ and
$h^{2,1}=115$. As explained in \cite{VW:tor}, $X^\sharp$ may be
written as a double cover of $(\P^1)^3$ and may deformed into $X^\prime$.
  \item[$X$] Let $f_{a_1a_2a_3}$ represent a generic polynomial of
weight $(a_1,a_2,a_3)$ in the homogeneous coordinates of
$(\P^1)^3$. Let W be a symmetric matrix of the form
\begin{equation}
  W=\left(\begin{array}{ccc} f_{400}&f_{220}&f_{202}\\
	f_{220}&f_{040}&f_{022}\\
	f_{202}&f_{022}&f_{004}\end{array}\right).
\end{equation}
$K$ is the hypersurface defined by $\det W=0$. $X$ is the double cover
of $(\P^1)^3$ branched over $K$. $K$, and therefore $X$, have 64
nodes.
  \item[$\bX$] The space $X$ may be blown-up to a smooth manifold
$\bX$ by replacing each of the 64 nodes by an exceptional divisor
$\P^1\times\P^1$. As before $\bX$ is not a \CY\ manifold.
\end{itemize}
These spaces are thus related as follows
\def\mpr{\mapright{\hbox{\scriptsize def}}}
\def\mpd{\mapdown{\hbox{\scriptsize blow-up}}}
\begin{equation}
\begin{array}{ccccc}
  X^\sharp&\mpr&X&\mpr&X^\prime\\
  \mpd&&\mpd&&\\
  Y&&\bX&&
\end{array}
\end{equation}
where ``def'' refers to a deformation of complex structure.

String theory on $X^\sharp$ is understood from orbifold theory. Since
$H_2(G)\cong\Z_2$ there are two possible theories depending on one's
choice of the ``discrete torsion'' 2-cocycle \cite{Vafa:tor}. With a
trivial 2-cocycle  one recovers the usual blow-up picture as expected
\cite{me:orb2}. That is, the chiral ring corresponds to the cohomology
of $Y$. Thus $Y$ may be taken to be the geometrical interpretation of
a conformal field marginally perturbed from that of the orbifold
$X^\sharp$ with trivial 2-cocycle.

When the nontrivial 2-cocycle is chosen, one obtains a chiral ring
mirror to that with a trivial 2-cocycle. That is, $h^{1,1}=3$ and
$h^{2,1}=51$. These numbers precisely agree with the degrees of freedom
of $X$. $X$ has 3 deformations of its
``K\"ahler form,'' that is,
the sizes of the three $\P^1$'s may be varied. ($X$ itself is singular
so it doesn't really have a K\"ahler form as such.)
Varying $W$ gives $K$ 30 deformations of complex structure but this
does not actually account for all the deformations of $K$ which
preserve the 64 nodes. Since $X^\prime$ has 115 deformations there
will be $115-64=51$ deformations of $K$ maintaining 64 nodes and thus
51 deformations of $\bX$. Thus there are 51 deformations of complex
structure for $X$ for our purposes.

As the reader may have guessed by now, the group $H_2(\bX)$ contains a
$\Z_2$ torsion part. Thus by analogous reasoning to the previous
example, if we set $\alpha=1$ for the A-model on $X$, we obstruct the
deformations taking $X$ into $X^\prime$. This then appears to give the
correct geometrical picture for allowing $X$ to be regarded as the
geometrical interpretation of a conformal field theory marginally
perturbed from that of the orbifold $X^\sharp$ with nontrivial
2-cocycle. Note however that in addition to just knowing the classical
geometry of $X$, we also need to put $\alpha=1$ to stop $X^\prime$
{}from providing the geometrical interpretation.


\section{Discussion}	\label{s:conc}

We have observed that conformal field theory, or topological field
theory, on a target space with nodes may have degrees of freedom which
are ``hidden away'' in the nodes. By perturbing by an irrelevant
operator we have been able to probe the secrets of these nodes to
discover the Brauer group at work.

It is important to realize in the above description that the hidden
degrees of freedom cannot be expressed as some local property of each
of the nodes. The appearance of torsion in $H_2(\bX)$ is a global
property --- many nodes in just the right place are required to
produce the element of the Brauer group on blowing up. This
demonstrates further some of the peculiar properties of the stringy
description of space.

A question we have not addressed is that of the existence of a
conformal field theory associated to some singular target space
$X$. Given a smooth \CY\ manifold near the large radius limit we may
assume the existence of a conformal field theory approximated by the
non-linear \sm\ with Ricci-flat target space. In the case of a
singular target space however some of the correlation functions of the
supposed conformal field theory may contain divergences.

Consider the conformal field theory on the manifold $X^\prime$ in the
example in either section \ref{ss:eg1} or \ref{ss:eg2} and consider
the process of deforming the target space continuously to $X$. Such a
degeneration of complex structure leads to infinities in the chiral
ring. That is, the B-model on $X^\prime$ appears bad in the limit
$X^\prime\to X$. This appears to rule out a good conformal field
theory corresponding to the A-model
with $\alpha=0$. For the case $\alpha=1$ we have removed
precisely the offending fields from the B-model causing the
divergences. Thus the case $\alpha=1$ contains no infinities and
may describe a good conformal field theory.

This agrees with the analysis of \cite{VW:tor} where there are only
two choices of orbifold theories on $X^\sharp$. One consists of the
theory which may be blown up to $Y$. The other is the theory which is
deformed to $X$ with $\alpha=1$. There is no third possibility of a
theory which may be deformed to $X$ with $\alpha=0$ since such a
theory would be a limit of $X^\prime$ and, as such, contain
divergences. Thus although we have introduced the Brauer group as an
extra parameter in the space of A-models, it would appear that to
obtain a finite conformal field theory on a singular space such as
$X$, one is forced to rule out the choice $\alpha=0$. Presumably it is
only on smooth manifolds that one is really free to choose $\alpha$.

The example in section \ref{ss:eg2} appears to show a link between
elements of the Brauer group and nontrivial 2-cocycles for
orbifolds. The imposition of a (non)trivial 2-cocycle for the orbifold
$X^\sharp$ appears to match the (non)trivial choice for $\alpha$ for
the theory on $X$. Can we therefore claim to have a complete geometrical
understanding of these 2-cocycles? Unfortunately the picture is not
complete. It is not possible to blow-up $X^\sharp$ to obtain some
manifold with nontrivial Brauer group. We must deform $X^\sharp$ into
$X$ before blowing up for our construction to work.

The desired theorem for a general case might appear along the lines as
follows. {\em
Given an orbifold $X^\sharp=V/G$, for finite group $G$, there exists some
$X$ obtained by a
deformation of complex structure of $X^\sharp$ such that the blow-up,
$\bX$, of $X$ satisfies $\Tors(H_2(\bX))\cong H_2(G)$.} In light of the
example of \cite{AM:suff} we must also exclude the trivial case where
$X^\sharp$ is a manifold. It is not at all clear that this conjecture is
true and it is certainly worthy of further study.

It was observed in \cite{VW:tor} that $X$ and $Y$ from section
\ref{ss:eg2} are a mirror pair. This is a fact that we have not used
yet. Since $Y$ may be written as a complete intersection in a
toric variety one should be able to use to method of
\cite{Boris:m,AG:gmi} to construct its mirror. This example is very
similar to that studied in section 3.4 of \cite{AG:gmi}. The result is
that the mirror of $Y$ is a hybrid model which is a ``trivial'' (i.e.,
quadratic) Landau-Ginzburg theory in $\C^6/\Z_2$ fibred over
$(\P^1)^3$. The superpotential of this Landau-Ginzburg theory
``degenerates'' (i.e, some directions become massless) over a subspace
of $(\P^1)^3$. This subspace appears in the form of $K$ in section
\ref{ss:eg2}. That is to say, the description of the mirror of $Y$ in
the language of \cite{AG:gmi} is precisely $X$ except that ``double
cover'' is replaced by ``Landau-Ginzburg fibration'' and ``branched
over'' is replaced by ``with superpotential degenerating over.'' This
is a curious point which should be pursued further.

\section*{Acknowledgements}

It is a pleasure to thank R.~Plesser for helpful
comments regarding section \ref{s:ncrit}, and B.~Greene for useful
conversations.
The help of M.~Gross was important for many points in the main text.
The work of P.S.A. is partially supported by a grant from the National
Science Foundation,
and that of D.R.M. by NSF grant DMS-9401447.

\setcounter{equation}{0}
\def\theequation{A.\arabic{equation}}
\section*{Appendix: Some calculations of Brauer groups}

\begin{center}
{\bf Mark Gross\footnote{Supported in part by
NSF grant DMS-9400873}}\\*
{\sl Department of Mathematics\\*
Cornell University\\*
Ithaca, NY 14853}\\*
{\tt mgross@math.cornell.edu}
\end{center}

\def\Brp{\mathop{{\rm Br}^\prime}}
\def\Br{\mathop{\rm Br}}
\def\Gm{{\Bbb G}_m}
\def\exact#1#2#3{0\rightarrow#1\rightarrow#2\rightarrow#3\rightarrow0}
\def\Pic{\mathop{\rm Pic}}
\def\qz{{\Q}/{\Z}}
\def\im{\mathop{\rm im}}
\def\Pone{\P^1}
\def\Pthree{\P^3}
\def\Gr{\mathop{\rm Gr}}
\def\Cl{\mathop{\rm Cl}}
\def\E{{\cal E}}
\def\I{{\cal I}}
\def\coker{\mathop{\rm coker}}
\def\ftn#1{{\hbox{\scriptsize #1}}}
\def\Pp{{\bf P}}

In this appendix, all cohomology groups will be defined using the
\'etale topology unless otherwise noted. See \cite{Milne:} for a basic
reference for the \'etale topology and \'etale cohomology.  By the
{\it Brauer group} of an algebraic variety $X$, we mean the
cohomological Brauer group, $\Brp(X)=H^2(X,\Gm)$, where $\Gm$ is the
sheaf of units in $\O_X$. See \cite{Grot:} for an introduction to
Brauer groups of varieties.  We will assume $X$ is defined over $\C$,
or any algebraically closed field of characteristic zero.

There is an exact sequence
$$\exact{\Pic(X)\otimes_{\Z} \qz}{H^2(X,\qz)}{\Brp(X)}.$$ (See
\cite{Grot:}, II Thm 3.1.) If $X$ is a non-singular variety over the
complex numbers, then $H^2(X,\qz)$ coincides with the singular
cohomology group $H^2_{\ftn{sing}}(X,\qz)$ in the usual topology. If
furthermore, $\Pic(X)\cong H^2_{\ftn{sing}}(X,\Z)$, as is the case if
$H^1(\O_X)=H^2(\O_X)=0$, then this exact sequence along with the
universal coefficient theorem shows that $\Brp(X)\cong
H^3_{\ftn{sing}}(X,\Z)_{\ftn{tors}}$.

We use the cohomological Brauer group rather than the description
``torsion in $H^3$'' because, for the second example below, we will
need some technical machinery which has already been set up using
\'etale cohomology in \cite{DoGr:}.  Furthermore, in the first example, we
will use the following interpretation for elements in the Brauer
group.

If $X$ is a variety, a Brauer-Severi variety over $X$ is a variety
$\Pp$ along with a map $f:\Pp\rightarrow X$ which is a
$\P^n$-bundle. (See \cite{Grot:}, I, \S 8 for details.)  Not all such
$\P^n$-bundles are projectivizations of vector bundles on $X$, and the
Brauer group gives obstructions for a Brauer-Severi variety to come
{}from a vector bundle. From the exact sequence
$$\exact{\Gm}{GL_{n+1}}{PGL_{n+1}}$$ we get, using suitably defined
cohomology groups, an exact sequence
$$H^1(X,GL_{n+1})\mapright{} H^1(X,PGL_{n+1}) \mapright{\delta_{n+1}}
H^2(X,\Gm).$$ Given a $\P^n$-bundle over $X$ is equivalent to giving a
class $\xi\in H^1(X,PGL_{n+1})$.  If $\delta_{n+1}(\xi)\not=0$, then
$\xi$ does not come from a rank $n+1$ vector bundle. Furthermore,
$\im\delta_{n+1}$ is annihilated by multiplication by $n+1$. (See
\cite{Grot:}, I, 1.4) Thus,
in particular, if $f:\Pp\rightarrow X$ is a $\Pone$-bundle which is not
the projectivization of a rank 2 vector bundle, then it gives rise to
a non-trivial 2-torsion element in $\Brp(X)$.

We now construct 2-torsion elements in the Brauer groups for the
threefolds mentioned in the main text.

{\it Construction 1.} Let $W\subseteq \Pthree\times\Pthree$ be a
generic hypersurface of bidegree $(2,2)$, and let $K\subseteq\Pthree$
be the discriminant locus of the fibration $p_1:W\rightarrow\Pthree$,
where $p_1$ is the projection onto the first factor. It is easy to see
that $K$ is an octic surface with 80 ordinary nodes. Let
$d:X\rightarrow\Pthree$ be the double cover of $\P^3$ branched over $K$,
and let $\pi:\tilde X\rightarrow X$ be the blowing-up of the 80 nodes
of $X$, so that $\tilde X$ is non-singular.

\proclaim Theorem 1. There is a non-trivial 2-torsion element in
$\Brp(\tilde X)$.

Proof: Let $U$ be the non-singular locus of $X$, so that $U=\tilde
X-\bigcup_{i=1}^{80} E_i$ where the $E_i$ are the exceptional divisors
obtained from blowing up the singular points of $X$. Each $E_i$ is a
non-singular quadric surface. By \cite{Grot:}, III 6.2, there is an exact
sequence
$$0\rightarrow{\Brp(\tilde
X)}\rightarrow{\Brp(U)}\rightarrow{\bigoplus_{i=1}^{80}
H^1(E_i,\qz)=0},$$ so $\Brp(\tilde X)=\Brp(U)$.  Now each point $x\in
U$ corresponds to a choice of a ruling of the non-singular quadric or
quadric cone $p_1^{-1}(d(x))$. Let $\Pp\subseteq \Gr(2,4)\times U$ be
the variety such that $\Pp_x$ parametrizes the lines in the
corresponding ruling of $p_1^{-1}(d(x))$, so that $f:\Pp\rightarrow U$
is a $\Pone$-bundle. Let $l_x\subseteq W$ be the line corresponding to
a point $x\in\Pp$.

{\it Claim:} $f$ does not have a rational section, i.e. a rational map
$\sigma: U\rightarrow \Pp$ with $f\circ\sigma$ the identity wherever
$\sigma$ is defined.

Proof: Suppose that $f$ has a rational section
$\sigma:U\rightarrow\Pp$. Let $D\subseteq W$ be defined to be the
Zariski closure of the set
\[\eqalign{
\{l_{\sigma(x_1)}\cap l_{\sigma(x_2)}|&\hbox{$x_1,x_2\in U$ are any
distinct points on which}\cr
  &\qquad\hbox{$\sigma$ is defined such that
$d(x_1)=d(x_2)$}\}.\cr}
\]
If $d(x_1)=d(x_2)$ then $l_{\sigma(x_1)}$ and
$l_{\sigma(x_2)}$ are lines in distinct rulings of $p_1^{-1}(d(x_1))$,
so the intersection consists of one point.  Thus the projection
$D\rightarrow\Pthree$ is generically one to one, and so the cup
product of the cohomology class $[D]$ of $D$ in $H^4_{\ftn{sing}}(W,\Z)$
with the cohomology class of a fibre of $p_1$ is one. But since $W$ is
ample in $\Pthree\times\Pthree$, by the Lefschetz hyperplane theorem,
$H^4_{\ftn{sing}}(W,\Z)\cong H^4_{\ftn{sing}}(\Pthree\times
\Pthree,\Z)$ and so the intersection of every cohomology class in
$H^4_{\ftn{sing}}
(W,\Z)$ with a fibre of $p_1$ is always even.  This is a
contradiction, proving the claim. $\bullet$

Now if $f$ were the projectivization of a rank 2 vector bundle $\E$ on
$U$, a section of $\E$ would yield a rational section of $f$. Thus
$f:\Pp\rightarrow U$ gives rise to a non-trivial 2-torsion element in
$\Brp(U)\cong \Brp(\tilde X)$.  $\bullet$

{\it Construction 2.} Let $P=\Pone\times\Pone\times\Pone$ with
trihomogeneous coordinates $([p_0,p_1],[p_2,p_3],\allowbreak [p_4,p_5
])$. We denote by
$\O_P(a,b,c)$ the line bundle of tridegree $(a,b,c)$.  Let $M$ be a
symmetric matrix
$$M=\pmatrix{f_{400}&f_{220}&f_{202}\cr f_{220}&f_{040}&f_{022}\cr
f_{202}&f_{022}&f_{004}\cr}$$ with the tridegrees of the forms
indicated by the subscripts. For general choice of $M$, $M$ is rank
$\le 2$ on a surface $K\subseteq P$ of tridegree $(4,4,4)$ whose
singular locus is the locus where $M$ is rank 1: from this, it is an
easy Chern class calculation using \cite{HarTu:} to see that $K$ has 64
nodes.

Now consider the map $s:P\rightarrow P$ defined by
$s([p_0,p_1],[p_2,p_3],[p_4,p_5])=
([p_0^2,p_1^2],\allowbreak [p_2,p_3],\allowbreak [p_4,p_5 ])$, so that
$s$ is a double cover.  Consider a general matrix
$$M_s=\pmatrix{f_{200}&f_{120}&f_{102}\cr f_{120}&f_{040}&f_{022}\cr
f_{102}&f_{022}&f_{004}\cr}.$$ $\det M_s$ vanishes on a surface $K_s$ of
tridegree $(2,4,4)$, which for general $M_s$ has 32 singular
points. Now $s^{-1}(K_s)$ is a surface $K$ of the type described above,
but the matrix $M=s^*M_s$ determining it may not be general.
Nevertheless, if $K_s$ is general, $K$ will have 64 nodes. If $X$ and
$X_s$ are the double covers of $P$ branched over $K$ and $K_s$
respectively, $\tilde X$ and $\tilde X_s$ the blow-ups of the nodes,
then it is clear that $\tilde X$ can be deformed smoothly to the
blow-up of the double cover branched over a surface determined by a
general matrix $M$. The Brauer group is a topological invariant, and
so showing $\Brp(\tilde X)$ contains a 2-torsion element for this
special $M$ will show it contains a 2-torsion element for general $M$.

Consider the map $f_s:X_s\rightarrow\Pone\times\Pone$ which is the
composition of the maps $X_s\rightarrow P$ and
$P\rightarrow\Pone\times\Pone$ given by projection onto the second and
third $\Pone$'s. $f_s$ is a conic bundle. We also define $f:X\rightarrow
\Pone\times\Pone$ similarly, so that $f$ is an elliptic fibration.
The following Lemma summarizes
the geometric results about $X$ and $X_s$ we will need.

\proclaim Lemma 2.
\begin{itemize}
\item[(1)] The discriminant locus $\Delta$ of $f_s$ consists
of two curves $\Delta_1$ and $\Delta_2$, each of type $(4,4)$ on
$\Pone\times\Pone$, meeting transversally at 32 points, and each fibre
of $f_s$ over $\Delta$ is a union of two $\Pone$'s.
\item[(2)] The Cartier divisor class group and the Weil divisor class group of
$X$ coincide, and $\Pic X\cong\Z^{\oplus 3}$, generated by
$p_i^*\O_{\Pone}(1)$, $1\le i\le 3$, where $p_i:X\rightarrow\Pone$ is
the projection onto the $i$th component of $P$.
\item[(3)] There is a non-singular threefold $V$ birationally equivalent
to $X$, and a map $g:V\rightarrow B$ birationally equivalent
to $f:X\rightarrow\Pone\times\Pone$, where $B$ is the blow-up of
$\Pone\times\Pone$ at the points of $\Delta_1\cap\Delta_2$. Furthermore
\begin{itemize}
\item[(a)] $g$ is flat.
\item[(b)] If we also denote by $\Delta_1$ and $\Delta_2$ the proper
transforms of these two curves on $B$, $g^{-1}(\Delta_1)$ and
$g^{-1}(\Delta_2)$ are irreducible divisors, and a fibre of $g$ over a
general point of $\Delta_1$ or $\Delta_2$ consists of a union of
$\Pone$'s meeting at two points.
\end{itemize}
\end{itemize}

Proof: (1) It is easy to see that for a general choice of $M_s$, the
projection $p:K_s\rightarrow\Pone\times\Pone$ is finite. $p$ is a
double cover, and the branch locus of $p$ is the discriminant locus
$\Delta$ of $f_s$. Furthermore, the singular points of $\Delta$ are
precisely the images of the singular points of $K_s$. Since $K_s$ has 32
nodes, $\Delta$ has 32 nodes.

Consider the curve $\Delta_1\subseteq\Pone\times\Pone$ defined by
$f_{040}f_{004} -f_{022}^2=0$. This is a curve of bidegree
$(4,4)$. Over this curve, $\det M_s$ reduces to
$$-f_{120}^2f_{004}+2f_{120}f_{102}f_{022}-f_{102}^2f_{040}.$$ If we
consider this as a quadratic expression in the variables $f_{120}$ and
$f_{102}$, its discriminant is $-4(f_{040}f_{004}-f_{022}^2)$, which
is zero over $\Delta_1$. Thus $K_s$ is branched over $\Delta_1$, and
$\Delta_1\subseteq\Delta$. It is easy to see that $\Delta$ is of
bidegree $(8,8)$ on $\Pone\times\Pone$, and thus
$\Delta=\Delta_1\cup\Delta_2$ with $\Delta_2$ of bidegree $(4,4)$.
Since $\Delta$ has 32 nodes, this leaves no choice but for $\Delta_1$
and $\Delta_2$ to be non-singular curves meeting transversally.

(2) Let $\Cl(X)$ denote the Weil divisor class group of $X$. The {\it
defect} of $X$ is defined as ${\rm rk}(\Cl(X)/\Pic(X))$, and this can
be computed
via the methods of \cite{Clem:}, \S 3 to be
$$\dim H^0(\I_{Z/P}(4,4,4))-\dim H^0(\O_P(4,4,4))+\hbox{\# of nodes of
$K$}$$ where $Z$ is the singular locus of $K$ and $\I_{Z/P}$ is the
ideal sheaf of $Z$ in $P$.

$Z$ is defined by the $2\times 2$ minors of the symmetric matrix
$M$. There are six such distinct minors, and using them, one obtains a
three step resolution of $\I_{Z/P}$, by direct sums of line bundles on
$P$. (One can do this by hand or very quickly using Macaulay \cite{Macly:}.)
{}From this one computes that $\dim H^0(\I_{Z/P}(4,4,4))=61$. We omit
the details. This then gives that the defect is zero.

Now $\Pic(X)\cong{\Z}^{\oplus 3}$ since $K$ is an ample divisor in
$P$, and the local class group of a node is torsion free, so
$\Cl(X)/\Pic(X)$ is torsion free and rank $0$. We conclude that
$\Cl(X)\cong\Pic(X)$.

(3) $X$ is a double cover of $X_s$ branched over a non-singular surface
$S$ which is contained in the non-singular part of $X_s$. If $p\in
\Delta_1\cap\Delta_2$, then $f^{-1}(p)=l_1\cup l_2$ with $l_1$ and
$l_2$ being $\Pone$'s intersecting at a node of $X_s$. Blow-up the node
and then the proper transforms of $l_1$ and $l_2$. Doing this for all
$p\in\Delta_1\cap\Delta_2$, we obtain a non-singular threefold $V_s$
with a flat morphism $V_s\rightarrow B$. (Equivalently, $V_s$ is
obtained by blowing up the singular locus of
$X_s\times_{\Pone\times\Pone} B$.) Let $S'$ be the proper transform of
$S$ in $V_s$. Since $S$ intersects $l_1$ and $l_2$ transversally, $S'$
is non-singular. Let $V$ be the double cover of $V_s$ branched along
$S$, and $g:V\rightarrow B$ the composition of $V\rightarrow V_s$ and
$V_s\rightarrow B$.  It is then clear from the construction that $g$ is
flat.  For (b), observe that if $g^{-1}(\Delta_i)$ was not
irreducible, then the Weil divisor class group of $X$ would be larger
than (2) permits. The last statement follows from the above
description of $V$.  $\bullet$

\proclaim Theorem 3. $\Brp(\tilde X)$ contains a non-trivial 2-torsion element.

Proof: The Brauer group is a birational invariant (\cite{Grot:}, III, Theorem
7.4), so it will be enough to show that $\Brp(V)$ contains 2-torsion.

We will follow the notation of \cite{DoGr:}. Let $\eta$ be the generic point
of $B$, $i:\eta\rightarrow B$ the inclusion, $P_{V/B}=R^1g_*\Gm$, and
$\E=\ker(P_{V/B}
\rightarrow i_*i^*P_{V/B})$. By Lemma 2, (3), $g:V\rightarrow B$ is
what is called a
good model in \cite{DoGr:}, Definition 1.1, so we can apply the results of
\cite{DoGr:}, \S 1.

Let $D_i=g^{-1}(\Delta_i)$, $\tilde D_i$ be the normalization of
$D_i$, and let $\tilde D_i\rightarrow
\tilde\Delta_i\rightarrow\Delta_i$ be the Stein factorization.  By
Lemma 2, (1), $\tilde\Delta_i\rightarrow\Delta_i$ is an unramified
double cover, as one obtains the same double covering using the map
$f_s$. By \cite{DoGr:}, Proposition 1.13,
$$\eqalign{H^1(B,\E)&=\bigoplus_{i=1}^2\ker(H^1(\Delta_i,\qz)\rightarrow
H^1(\tilde
\Delta_i,\qz))\cr
&=(\Z/2\Z)^{\oplus 2}.\cr}$$ By the exact sequence
$$\exact{\E}{P_{V/B}}{i_*i^*P_{V/B}}$$ (surjectivity on the right
follows from \cite{DoGr:}, Proposition 1.10) we obtain a sequence
$$H^0(B,P_{V/B})\mapright{\alpha} H^0(B,i_*i^*P_{V/B})\mapright{}
H^1(B,\E)\mapright{} H^1(B,P_{V/B}).$$

We also have an exact sequence of sheaves on $\eta$:
$$0\mapright{} A\mapright{} i^*P_{V/B}\mapright{d} \Z\mapright{}
0$$ where $d$ is the degree map and $A$ is the Jacobian of
$V_{\eta}=X_{\eta}$. First $H^0(A)=0$: a non-trivial degree zero line
bundle on $X_{\eta}$ would extend to give a divisor on $X$ not allowed
by Lemma 2, (2), and so $H^0(\eta,i^*P_{V/B})\subseteq\Z$. Thus
$\coker(\alpha)$ is a cyclic group, and since it injects into
$H^1(B,\E)$, we must have
$$\coker(\alpha)=\hbox{$0$ or $\Z/2\Z$,}$$ and thus
$$\im(H^1(B,\E)\rightarrow H^1(B,P_{V/B}))=\hbox{$\Z/2\Z$ or
$(\Z/2\Z)^{\oplus 2}$.}$$ In fact, it is the first case which
occurs, but since that does not matter to us, we do not prove this
here.  Finally, we have an exact sequence (\cite{DoGr:}, 1.5)
$$0=\Brp(B)\rightarrow\Brp(V)\rightarrow H^1(B,P_{V/B})\rightarrow
H^3(B,\Gm)=0,$$ with the left and right terms being zero since $B$ is
a rational surface, so $\Brp(V)$ contains a two-torsion
element. $\bullet$

{\it Remark 4.} The above computation obscures the actual source of
the 2-torsion, which can be seen in the following manner: Using the
results of \cite{AM:tor}, $H^4_{\ftn{sing}}(V_s,\Z)$ has
2-torsion, generated by
$l_0-l_1$, where $l_0$ and $l_1$ are the two components of a fibre of
$g$ over a point in $\Delta_1$ or $\Delta_2$.  ($V_s$ is as in the
proof of Lemma 2, (3).)  This cycle then lifts to a difference of two
rational curves in $V$, generating the 2-torsion in
$H^4_{\ftn{sing}}(V,\Z)$ we have produced above. However, this requires
a more detailed analysis of the geometry of $V$. Furthermore, the
method of proof of Theorem 4 is more suitable an approach for some
other examples.


\end{document}